\documentclass[dvips]{article}
\usepackage{graphicx}
\newcommand{\doublespacing}{\let\CS=\@currsize\renewcommand{\baselinesstrech}
{2.0}\tiny\CS}

\begin{document}

\newcommand{\bd}{\begin{document}}
\newcommand{\ed}{\end{document}}
\newcommand{\bc}{\begin{center}}
\newcommand{\ec}{\end{center}}
\newcommand{\bfr}{\begin{flushright}}
\newcommand{\efr}{\end{flushright}}
\newcommand{\lt}{\left}
\newcommand{\rt}{\right}
\newcommand{\vs}{\vspace}
\newcommand{\hs}{\hspace}
\newcommand{\beq}{\begin{equation}}
\newcommand{\eeq}{\end{equation}}
\newcommand{\lb}{\linebreak}
\newcommand{\pb}{\pagebreak}
\newcommand{\mb}{\makebox}
\newcommand{\fb}{\framebox}
\newcommand{\mc}{\multicolumn}
\newcommand{\ben}{\begin{enumerate}}
\newcommand{\een}{\end{enumerate}}
\newcommand{\bit}{\begin{itemize}}
\newcommand{\eit}{\end{itemize}}
\newcommand{\ol}{\overline}
\newcommand{\un}{\underline}
\newcommand{\lefq}{\lefteqn}
\newcommand{\ba}{\begin{array}}
\newcommand{\ea}{\end{array}}
\newcommand{\beqa}{\begin{eqnarray}}
\newcommand{\eeqa}{\end{eqnarray}}
\newcommand{\beqas}{\begin{eqnarray*}}
\newcommand{\eeqas}{\end{eqnarray*}}
\newcommand{\bfg}{\begin{figure}}
\newcommand{\efg}{\end{figure}}
\newcommand{\bds}{\begin{displaymath}}
\newcommand{\eds}{\end{displaymath}}
\newcommand{\btb}{\begin{tabbing}}
\newcommand{\etb}{\end{tabbing}}
\newcommand{\para}{\parallel}
\newcommand{\pad}{\partial}
\newcommand{\nn}{\nonumber}
\newcommand{\la}{\leftarrow}
\newcommand{\ra}{\rightarrow}
\newcommand{\lgla}{\longleftarrow}
\newcommand{\lgra}{\longrightarrow}
\newcommand{\La}{\Leftarrow}\newcommand{\Ra}{\Rightarrow}
\newcommand{\Lra}{\Leftrightarrow}
\newcommand{\Lgla}{\Longleftarrow}
\newcommand{\Lgra}{\Longrightarrow}
\newcommand{\bm}{\boldmath}
\newcommand{\lan}{\langle}
\newcommand{\ran}{\rangle}
\renewcommand{\a}{\alpha}
\renewcommand{\b}{\beta}
\newcommand{\g}{\gamma}
\newcommand{\G}{\Gamma}
\renewcommand{\d}{\delta}
\newcommand{\eps}{\epsilon}
\newcommand{\Th}{\Theta}
\newcommand{\s}{\sigma}
\newcommand{\lam}{\lambda}
\newcommand{\D}{\Delta}
\newcommand{\vare}{\varepsilon}
\newcommand{\pr}{\prime}
\newcommand{\ro}{\rho}
\newcommand{\nab}{\nabla}
\newcommand{\m}{\mu}
\newcommand{\n}{\nu}
\newcommand{\Sg}{\Sigma}
\newcommand{\p}{\pi}
\newcommand{\R}{I\!\!R}
\newcommand{\om}{\omega}
\newcommand{\Om}{\Omega}
\newcommand{\ze}{\zeta}
\newcommand{\vart}{\vartheta}
\newcommand{\tri}{\triangle}
\newcommand{\f}{\frac}
\newcommand{\iny}{\infty}
\newcommand{\pro}{\propto}
\bc {\huge \bf Exact solution of the Klein Gordon equation in the presence of a minimal length}
\ec
\vs{2cm}

\bc {\ T.K. Jana {\footnote{email : tapasisi@gmail.com} and P. Roy{\footnote{email : pinaki@isical.ac.in}}\\
Physics \& Applied Mathematics Unit \\
Indian Statistical Institute \\
Kolkata - 700 108, India.}} \ec \vspace{3.5cm}

\bc {\large {\un{Abstract}}} \ec
We obtain exact solutions of the $(1+1)$ dimensional Klein Gordon equation with linear vector and scalar potentials in the presence of a minimal length. Algebraic approach to the problem has also been studied.

\pb
\section{Introduction}
During the past few years, there have been growing interest in obtaining exact solutions of relativistic wave equations. In particular exact solutions of the Klein Gordon equation with various vector and scalar potentials have been obtained by a number of authors \cite{chen,castro}. In all these cases the models have been studied within the context of point particles.

On the other hand the concept of a minimal length has emerged from studies on quantum gravity \cite{garay}, perturbative string theory \cite{gross2}, black holes \cite{magg} etc. In such a scenario the standard Heisenberg uncertainty relation gets modified and this causes UV/IR mixing. As a consequence it is meaningful to study quantum mechanics in the presence of a minimal length \cite{kempf1}. In particular exact as well as approximate solutions of various quantum mechanical problems have been obtained in the presence of a minimal length \cite{kempf1,chang}. 

It may be noted that Klein Gordon equation which is solvable for  different vector and scalar potentials without minimal length may not be exactly solvable when considered in the presence of a minimal length. Here we shall consider $(1+1)$ dimensional Klein Gordon equation with (unequal) linear scalar and vector potentials in the presence of a minimal length and it will be shown that the problem admits exact analytical solutions. More precisely we shall treat the problem in two ways: first we obtain the solutions by solving the Schr\"odinger like equation in momentum space and secondly we shall obtain the solutions in a purely algebraic fashion by utilizing the shape invariance symmetry of the problem \cite{spector}. 

\section{Klein Gordon equation in the presence of minimal length}
In one dimensional quantum mechanics in the presence of a minimal length the canonical commutation relation between position and momentum becomes deformed. Out of the various deformed commutation relations we shall consider here the simplest one and it is given by \cite{kempf1}
\beq
[{\hat x},{\hat p}] = i\hbar(1+\b p^2)\label{cano}
\eeq
where $\b\geq 0$ is a small parameter. A representation of $\hat{x}$ and $\hat{p}$ satisfying (\ref{cano}) are given by \cite{kempf1}
\beq
{\hat x} = i\hbar[(1+\beta p^2)\f{\partial}{\partial p}+\gamma p],~~~~ {\hat p} = p \label{rep1}
\eeq
 
Also as a consequence of (\ref{cano}) the Heisenberg uncertainty relation gets modified and the generalized uncertainty relation reads
\beq
\Delta{\hat x}\Delta {\hat p} \geq \f{\hbar}{2}[1+\beta (\Delta {\hat p})^2]\label{uncer}
\eeq 
From (\ref{uncer}) it follows that there also exist a minimal length given by
\beq
(\Delta {\hat x})_{min} = \hbar \sqrt{\beta}
\eeq
In the space where position $({\hat x})$ and momentum $({\hat p})$ are given by (\ref{rep1}) the associated scalar product (which ensures hermiticity of $\hat{x}$ and $\hat{p}$) is defined by
\beq
\left<\phi(p)|\psi(p)\right> = \int \f{\phi^*(p)\psi(p)}{(1+\beta p^2)^{1-\f{\gamma}{\beta}}}~dp\label{scalar1}
\eeq

We note that the form of the $(1+1)$ dimensional Klein Gordon equation in the presence of a minimum length is similar to the standard one \cite{chen} except that the position and momentum are now given by (\ref{rep1}) : 
\beq
[c^2p^2 + (mc^2 + S(\hat x))^2]\psi = [E - V(\hat x)]^2\psi\label{kg1}
\eeq
We now choose the vector and the scalar potential to be of the form
\beq
V(\hat x) = \mu {\hat x},~~~~S(\hat x) = \lam{\hat x}\label{vs}
\eeq
where we have taken $\lam^2>\mu^2$ so as to avoid complex eigenvalues. Now using the representation (\ref{rep1}) the Klein Gordon equation (\ref{kg1}) can be written in momentum space as
\beq
\left[-f(p)\f{d^2}{dp^2} + g(p)\f{d}{dp} + h(p)\right]\psi = \eps\psi\label{kg2}
\eeq
where the functions $f(p), g(p)$ and $h(p)$ are given by
\beq
\ba{lcl}
f(p) &=& \displaystyle(1+\beta p^2)^2\\
g(p) &=& \displaystyle-2(1+\beta p^2)\left[p(\beta+\g)- i\f{(\mu E+mc^2\lam)}{\hbar(\lam^2-\mu^2)}\right]\\
h(p) &=& \displaystyle-p^2\left[\g(\beta+\g)-\f{c^2}{\hbar^2(\lam^2-\mu^2)}\right]+\f{2i\g p~(\mu E+mc^2\lam)}{\hbar(\lam^2-\mu^2)}\\
\eps &=& \g + \displaystyle\f{(E^2-m^2c^4)}{\hbar^2(\lam^2-\mu^2)}\\
\ea\label{eps}
\eeq

Eq.(\ref{kg2}) can be solved by performing a transformation involving a change of variable as well as wavefunction:
\beq
\psi(p) = \rho(p)\phi(p),~~~~q = \int \f{1}{\sqrt{f(p)}}~dp\label{t}
\eeq
where 
\beq
\rho(p) = e^{\int \chi(p)~dp},~~~~\chi(p) =  \f{f^\prime+2g}{4f}\label{rho}
\eeq
Under the above transformation Eq.(\ref{kg2}) can be written in the form of a Schr\"odinger equation
\beq
\left[-\f{d^2}{dq^2}+V(q)\right]\phi = e\phi\label{kg3}
\eeq
where the potential $V(q)$ and the energy $e$ are given by
\beq
\ba{lcl}
V(q) &=& \displaystyle \f{c^2}{\beta\hbar^2(\lam^2-\mu^2)}sec^2(\sqrt{\beta}q),~~~~-\f{\pi}{2\sqrt{\beta}}< q <\f{\pi}{2\sqrt{\beta}}\\
e &=& \displaystyle\f{(\lam E+\mu mc^2)^2}{\hbar^2(\lam^2-\mu^2)}+\f{c^2}{\hbar^2(\lam^2-\mu^2)\beta}
\ea\label{pot}
\eeq
The potential appearing above is a standard solvable potential whose energy and eigenfunctions (apart from a normalization factor) are given by \cite{khare}
\beq
e_n = (A+\sqrt{\beta}n)^2,~~~~A = \f{\sqrt{\beta}+\sqrt{\beta+\f{4c^2}{\hbar^2(\lam^2-\mu^2)\beta}}}{2},~~~~n=0,1,2,....\label{ener1}
\eeq
\beq
\phi_n(q) = (cos\sqrt{\beta}q)^{\f{A}{\sqrt{\beta}}}P_n^{(\f{A}{\sqrt{\beta}}-\f{1}{2},\f{A}{\sqrt{\beta}}-\f{1}{2})}(\sin\sqrt{\beta}q)
\eeq
Now using the relations (\ref{eps}) and (\ref{t}) we finally obtain
\beq
E_n = -\f{\mu mc^2}{\lam}+\f{\hbar(\lam^2-\mu^2)}{\lam}\sqrt{\beta(n^2+n+\f{1}{2})+\beta(n+\f{1}{2})\sqrt{1+\f{4c^2}{\hbar^2\beta^2(\lam^2-\mu^2)}}}\label{ener1}
\eeq
\beq
\psi_n(p) =\displaystyle e^{i\f{(mc^2\lam+E_n\mu)tan^{-1}(\sqrt{\beta}p)}{\hbar\sqrt{\beta}(\lam^2-\mu^2)}}\left(1+\beta p^2\right)^{-(\f{\g}{2\beta}+\f{A}{\sqrt{\beta}})}P_n^{(\f{A}{\sqrt{\beta}}-\f{1}{2},\f{A}{\sqrt{\beta}}-\f{1}{2})}(\sin\sqrt{\beta}q)
\eeq
where $P_n^{(a,b)}$ denotes Jacobi polynomials. It can be shown that the eigenfunctions are orthogonal with respect to the scalar product (\ref{scalar1}). We note that although (\ref{ener1}) is an exact result, nevertheless it is sometimes useful to separate the $\beta$ dependent contribution to the spectrum. So expanding (\ref{ener1}) in powers of $\beta$ we find
\beq
E_n \approx -\f{\mu mc^2}{\lam}+\sqrt{\f{\hbar c}{\lam^2}}~(\lam^2-\mu^2)^{\f{3}{4}}\sqrt{(2n+1)}+\sqrt{\f{\hbar^3}{4c\lam^2}}~\left(\lam^2-\mu^2\right)^{\f{5}{4}}\f{(n^2+n+\f{1}{2})}{\sqrt{2n+1}}~\beta+O(\beta^2)\label{ener2}
\eeq
where the first two terms are the standard $\beta$ independent contribution \cite{chen} while the rest depends on $\beta$.

\section{Algebraic approach}
In the last section we obtained exact solutions of the Klein Gordon oscillator. Since the problem is exactly solvable it is natural to examine it's underlying symmetry. Here we shall show that the problem has shape invariance symmetry and use the approach suggested in \cite{spector} to obtain exact solutions in a purely algebraic fashion. We recall that two Hamiltonians $H_1(\lam)$ and $H_2(\lam)$ are said to be shape invariant if \cite{khare}
\beq
H_2(\lam_1) = H_1(\lam_2) + R(\lam_1)
\eeq
where $\lam_1$ is a set of parameters, $\lam_2$ is a function of $\lam_1$ and $R$ is a function independent of $p$. Then it can be shown that the spectrum is given by
\beq
E_n = \sum_{i=1}^n R(\lam_i)\label{ener2}
\eeq

To utilize the shape invariance property we shall now try to factorize the following Hamiltonian in the form 
\beq
H = -f(p)\f{d^2}{dp^2} + g(p)\f{d}{dp} + h(p) +c_1 = CB\label{cb}
\eeq
where $c_1$ is a constant (to be determined later) and the operators $B$ and $C$ are not necessarily self adjoint. \footnote[1] {Note that one can prove shape invariance of Eq.(\ref{kg2}) using shape invariance property of the potential (\ref{pot}) and the inverse transformation of (\ref{t}).} We consider these operators to be of the form 
\beq
\ba{lcl}
B &=& F(p) \f{d}{dp} + W(p) + \Omega (p)\\
C &=& -F(p) \f{d}{dp} + W(p) - \Omega (p)\label{bc}
\ea
\eeq 
From (\ref{bc}) it follows that
\beq
CB = -F^2(p)\f{d^2}{dp^2} - F(p)[F^\prime(p)+2\Omega(p)]\f{d}{dp} - F(p)[W^\prime(p)+\Omega^\prime(p)] + W^2(p) - \Omega^2(p)\label{cb1}
\eeq
Comparing (\ref{cb}) and (\ref{cb1}) we get $F(p),g(p)$ and an equation for $W(p)$:
\beq
\ba{lcl}
F(p) &=& (1+\beta p^2)\\
\Omega(p) &=& \displaystyle\g p - \f{i(\mu E+mc^2\lam)}{\hbar(\lam^2-\mu^2)}\\
\ea
\eeq

\beq
(1+\beta p^2)\f{dW}{dp} + W^2(p) = \f{c^2p^2}{\hbar^2(\lam^2-\mu^2)}+\g -\f{(\mu E+mc^2\lam)^2}{\hbar^2(\lam^2-\mu^2)^2}+c_1\label{W}
\eeq
To solve this equation we now consider an ansatz for $W(p)$:
\beq
W(p)= c_2p
\eeq
It can be shown that the Ricatti equation (\ref{W}) is satisfied if
\beq
\ba{lcl}
c_1 &=& \displaystyle\f{(\mu E+mc^2)^2}{\hbar^2(\lam^2-\mu^2)^2}-c_2-\g\\
c_2 &=& \displaystyle\f{\beta+\sqrt{\beta^2+\f{4c^2}{\hbar^2(\lam^2-\mu^2)}}}{2}~p\label{c}
\ea
\eeq
Then after some calculations it can be shown that
\beq
B(\lam_1)C(\lam_1) = C(\lam_2)B(\lam_2) + R(\lam_1)\label{si}
\eeq
where we have taken 
\beq
\lam_1=c_2,~~~~\lam_i = c_2+(i-1)\beta,~~~~R(\lam_i) = 2\lam_i+\beta
\eeq
Now iterating (\ref{si}) it follows that
\beq
\eps+c_1 = \sum_{i=0}^n R(\lam_i)
\eeq
Finally using (\ref{eps}) and (\ref{c}) one obtains $E_n$ and they are the same as in (\ref{ener1}). 
\section{Discussion}
Here we have obtained exact solutions of the Klein Gordon equation with linear vector and scalar potentials. We feel it would be interesting to search for other vector and scalar potentials for which exact or approximate solutions of the Klein Gordon equation can be obtained. We would like to mention that apart from solving the Klein Gordon equation we have also exploited the shape invariance symmetry of the problem to obtain the spectrum.  We feel it would be interesting to investigate other symmetries e.g, Lie algebraic symmetry of this class of problems whenever the transformation (\ref{t}) is invertible. It may also be noted here that we have obtained the trigonometric potential (\ref{pot}) as a consequence of the relation (\ref{cano}) and the choice of $S(\hat{x}),V(\hat{x})$. However this is not the only deformation of the canonical commutation relation involving $\hat{x}$ and $\hat{p}$. It may be interesting to search for some other choice of $[\hat{x},\hat{p}], S(\hat{x})$ and $V(\hat{x})$ which may eventually lead to other exactly solvable potentials e.g, hyperbolic P\"oschl Teller potential. Finally we would like to mention that higher dimensional analogue of the system considered here may have applications in phenomenology e.g, in the study of meson spectrum \cite{kang}.

\ed